\documentclass[twocolumn]{aa}
\usepackage{epsfig}
\usepackage{graphics}
\usepackage{color}
\usepackage{graphicx}
\usepackage[T1]{fontenc}
\usepackage{amsmath,amssymb}

\newcommand\bvec[1]{\boldsymbol{#1}}

\def\deg~{\ifmmode^\circ _\cdot\else$^\circ _ \cdot$\fi }    
\def\degg{\ifmmode^\circ \else$^\circ $\fi } 
\def\prima{^{\prime}}

{ \begin{list}%
        {$\bullet$}%
        {\setlength{\labelwidth}{20pt}%
         \setlength{\leftmargin}{25pt}%
         \setlength{\itemsep}{\parsep}}}%
{ \end{list} }

\begin{document}
   \title{On the wavelet analysis of CMB time ordered data: application to Archeops.}


   \author{Juan Francisco Mac\'{\i}as-P\'erez
          \inst{1}
          \and
           Alexandre Bourrachot \inst{2}
    }
     
   \offprints{macias@lpsc.in2p3.fr}

   \institute{Laboratoire de Physique Subatomique et Cosmologie, \\
              53 avenue des Martyrs, 38026 Grenoble CEDEX, France
              \email{macias@lpsc.in2p3.fr}
         \and
            Laboratoire de l'Acc\'el\'erateur Lin\'eaire\\ 
            BP 34, Campus Orsay, 91898 Orsay Cedex, France
             \email{bourrach@lal.in2p3.fr}
             }

   \date{Received 00, Accepted 00}

   \abstract{ We present an alternative analysis of CMB time ordered
   data (TOD) using a wavelet-based representation of the data
   time-frequency plane. We demonstrate that the wavelet transform
   decorrelates $1/f$-type Gaussian stationary noise and permits a
   simple and functional description of locally stationary
   processes. In particular, this makes possible the generalization of
   the classical algorithms of map making and CMB power spectrum
   estimation to the case of locally stationary 1/f type noise. As an
   example, we present a wavelet based algorithm for the destriping of
   CMB-like maps. In addition, we describe a wavelet-based analysis of
   the Archeops data including time-frequency visualization, wavelet
   destriping and filtering of the TOD. These filtered data was used
   to produce polarized maps of Galactic dust diffuse
   emission. Finally, we describe the modeling of the non-stationarity
   on the Archeops noise for the estimation of the CMB power spectrum.
}
\authorrunning{Mac\'{\i}as-P\'erez, J.F. \& Bourrachot, A.}
\titlerunning{Wavelet analysis of time ordered data}

   \maketitle
%
\section{Introduction}
\label{introduction}

After their discovery by the COBE satellite \cite{cobepaper}, the Cosmic Microwave
Background (CMB) temperature anisotropies have become one of the most powerful observational tests
for cosmology. Indeed, recent measurements of their angular temperature power spectrum by ground based
experiments such as DASI (\cite{dasipaper}), CBI (\cite{cbipaper}) and VSA (\cite{vsapaper})
and by balloon-borne experiments
such as BOOMERang (\cite{boompaper}), MAXIMA (\cite{maxpaper}) and Archeops (\cite{benoit_params,tristram_cl} have provided strong
constraints on the main cosmological parameters such as the Hubble constant, $H_{0}$,
the Universe total density, $\Omega_{0}$, matter density, $\Omega_{m}$, 
and energy density $\Omega_{\Lambda}$, 
and the scalar $n_{s}$, and tensor $n_{t}$ spectral indexes associated to the power spectrum
of the primordial fluctuations. More recently, the satellite mission WMAP \footnote{Launched by NASA in July 2001. http://lambda.gsfc.nasa.gov/product/map}
(\cite{wmap})
has measured both the temperature CMB angular power spectrum and the
cross correlation between temperature and polarization (\cite{wmappol}) leading to high precision
determination of the cosmological parameters.  
In the near future, the PLANCK satellite mission \footnote{To be launched by ESA in 2007. 
http://sci.esa.int/science-e/www/area/index.cfm?fareaid=17}
will significantly improve the accuracy on the measurement of the angular power spectrum of the 
CMB temperature anisotropies and will also measure the CMB polarization
anisotropies. This will lead to a rather more selective test for discriminating competing
cosmological models and to constraints on the inflationary paradigm.  \\

The analysis of the PLANCK satellite data will require both faster and more accurate
algorithms than those currently available, to deal with such large data sets containing 
few $10^9$ samples, and to measure polarized fluctuation signals which are two or more order
of magnitude smaller than temperature fluctuations. One of the most important problems
in the analysis of CMB data is the characterization of the properties of the noise in the 
time ordered data (TOD) and their exploitation for optimal map-making and for the
reconstruction of the angular power spectrum of the CMB temperature and polarization anisotropies.
The extremely large size of the data sets makes computationally unfeasible
the direct inversion of the noise covariance matrix and even its storage (\cite{madcap}). 
This leads to the common use of Fourier techniques (\cite{mapcumba,master,mirage})
which are exclusively valid in the case of stationary and Gaussian noise for which the covariance
matrix is Toeplitz and it is diagonalized by the Fourier transform. In the
case of PLANCK, observations will take place over two years and therefore
it seems reasonable to expect that the TOD noise properties will change along
the mission. Further, residuals from badly removed parasitic noises will also contribute
to the non stationarity and non Gaussianity of the TOD noise.\\

New promising
Fourier methods (\cite{mapmakingitalian}) based on the concept of {\it locally stationary noise} 
have been presented to deal with the non stationarity problem. However, they
are not well adapted to the case of residuals from parasitic signals, and they reduce
significantly the number of data samples to be used in the analysis so that the overall
instrument sensitivity also decreases. In addition, these new methods requires an a priori
knowledge of the zones of locally stationarity on the data but no algorithm has been yet proposed
to address this issue. In this paper, we describe a new approach to deal with non stationarity
in CMB data sets. This approach is based on the simultaneous analysis
of the time and the frequency properties of noise using the wavelet and wavelet packet transforms. 
By contrast to Fourier based methods, non stationarity is naturally described by the
wavelet transform and no cutting of the data is needed to properly define locally
stationary processes. In addition, we discuss how wavelet techniques are very useful to identify
systematics in the TOD both visually and automatically. The application of these techniques
to the Archeops data is presented going from the simple visual inspection for classification 
of the
data to specific designed algorithms to remove low frequency components in the TOD and remaining
stripes on the Archeops maps.\\

Archeops \footnote{{\tt http://www.archeops.org}.} is a balloon borne
bolometer experiment that aimed at measuring the CMB temperature
anisotropy over large and small angular scales. It provided the first
determination of the $C_\ell$ spectrum from the COBE multipoles
(\cite{cobepaper}) to the first acoustic peak (\cite{benoit_cl,tristram_cl}) from
which it gave a precise determination of various cosmological
parameters, such as the total density of the Universe and the baryon
fraction (\cite{benoit_params}). Archeops was also designed as a test
bed for Planck--HFI and therefore shared the same technological
design: a Gregorian off--axis telescope with a $1.5\,$m primary
mirror, bolometers operating at 143, 217,
353 and $545$~GHz which are cooled down at $100\,$mK by a $^{3}$He$/^{4}$He dilution
designed to work at zero gravity and similar scanning strategy. 
Because of this Archeops data are expected to have common features with
Planck data with respect to random noise and systematics.
A detailed description of the instrument and its performances can be
found in (\cite{benoit_app}). \\

The plan for this paper is the following. Section 2 introduces the wavelet
and wavelet packet transforms. Section 3 discusses the properties of the
wavelet transform of stationary Gaussian process. Section 4 deals with non
stationary noise and in particular locally stationary noise. Section 5 
discusses a wavelet based destriping algorithm. Finally, section 6 describes
a wavelet analysis of the Archeops TOD.

\section{The wavelet and wavelet-packet transforms}
In this section we give a brief overview of the wavelet and wavelet packet transforms
which will be extensively used in the following sections. 

\subsection{The continuous wavelet transform}
\label{cwavetransform}
The term wavelet is used to refer to a set of orthonormal basis functions generated
by dilation and translation, 
$\psi_{a,b}(t) = a^{-\frac{1}{2}} \ \psi \left( \frac{t-b}{a}\right)$ 
of a compactly supported function, $\psi (t)$ where
$a$ and $b$ are called the scaling and translation factors. $\psi (t)$ is 
called the wavelet function or mother wavelet and it is assumed to satisfy
the admissibility condition
\begin{equation}
C_{\psi} = \int\limits_{-\infty}^{+\infty}\  \frac{{\left| \hat{\psi} (w)\right|}^2}{\left|w\right|}
\ dw < \infty 
\end{equation}
where $\hat{\psi} (w)$ is the Fourier transform of $\psi (t)$ \\
The continuous wavelet transform (CWT) of the function $g(t)$ is defined by
\begin{equation}
w_{a,b} = \left< f, \psi_{a,b} \right>= 
\int\limits_{-\infty}^{+\infty}{g(t) \ \psi_{a,b}(t) \ dt}.
\end{equation}
where  $w_{a,b}$ are called the wavelet coefficients associated to the function $g(t)$. 
We can also define the inverse wavelet transform so that
\begin{equation}
g(t) = \frac{1}{C_{\psi}}
\int\limits_{0}^{+\infty} \int\limits_{-\infty}^{+\infty}  \ 
w_{a,b} \ \psi_{a,b}(t) \ \frac{da \ db}{a^{2}}.
\end{equation}
The CWT constitutes a powerful tool to describe
simultaneously the time and frequency properties of a given function $g(t)$.
In this sense, it supersedes the standard Fourier transform within the limits
of the Heisenberg's uncertainty principle $\Delta f \ \Delta t \geq \frac{1}{2} \ \ $
where $\ \ f = 2 \pi w $. \\

\subsection{The Discrete Wavelet Transform}
\label{wavetransform}
The discrete wavelet transform, denoted DWT hereafter, of a discretely sampled time series,
$\left\lbrace X(t_{m}), \ \ m \ = 0, \ldots, N-1 \right\rbrace$
can be obtained from the CWT,  $w_{a,b}$, setting $a=2^{j-1}$ and $b = 2^{j-1} \ k_{j}$ for $j= 1,J$ and
$k_{j}=0, \ldots, 2^{j-1}$ where $N=2^{J_{max}}$. In other words, the DWT is obtained by restricting the CWT
to a 'dyadic partition' in time. For each scale $\tau_{j}=2^{j-1}$,
with $j=1,\ldots, J$, the DWT is composed of $N_{j} = \frac{N}{2^{j}}$ wavelet coefficients 
$\mathbf{W}_{j}$-. \\

Therefore, the DWT is an orthonormal transform $\mathcal{W}$ that takes a time series - TOD -
$\mathbf{X}= \left\lbrace X_{0}, X_{1} \ ,\ldots, \ X_{N-1}\right\rbrace^{T}$ and yields
a vector of $N$ DWT coefficients 
\begin{equation}
\label{eqdwtdefmatrix}
\mathbf{W} \equiv \mathcal{W}\mathbf{X} =
\left\lbrace \mathbf{W}_{1}^{T},\mathbf{W}_{2}^{T},\ldots,\mathbf{W}_{J}^{T},\mathbf{V}_{J}^{T}\right\rbrace.
\end{equation}
As above, the sub-vector $\mathbf{W}_{j}$ contains $N_{j} = \frac{N}{2^{j}}$ wavelet coefficients associated
with scale $\tau_{j} \equiv 2^{j-1}$, whereas $\mathbf{V}_{J}$ contains $\frac{N}{2^{J}}$ scaling 
coefficients
associated with scale $\lambda_{J} \equiv 2^{J}$. These scaling coefficients are needed to compensate
for the truncation of the DWT at level $J$.
As it will be discussed later on the section, they account 
for the smooth component of the time series which is left over by the wavelet coefficients.
Indeed, for $J=J_{max} \equiv \lg_{2} N$ then  $\mathbf{V}_{J_{max}}$ is composed by a single scaling 
coefficient which represents the mean value of the time series.
The orthonormality condition $\mathcal{W}^{T}\mathcal{W}= I_{N}$ implies that the inverse DWT is given by
\begin{equation}
\label{eqidwtdefmatrix}
\mathbf{X} \equiv \mathcal{W}^{T}\mathbf{W}=
\left\lbrace X_{0}, X_{1} \ ,\ldots, \ X_{N-1}\right\rbrace^{T}
\end{equation}

We can formally describe the DWT in terms of wavelet and scaling filters as follows. Assume
$\mathbf{h}_{1} = \left\lbrace h_{1,0},\ldots, h_{1,L-1},0,\ldots,0\right\rbrace^{T}$ to be
a vector of length $N$ whose first $L < N$ elements are the unit wavelet filter coefficients
for a given compactly supported wavelet (\cite{daubbook}) and with discrete Fourier transform (DFT)  
$\left\lbrace H_{1,k}, k=0,\ldots, N-1 \right\rbrace$.
Then $\mathbf{g}_{1} = \left\lbrace g_{1,0},\ldots, g_{1,L-1},0,\ldots,0\right\rbrace^{T}$ is a
vector of length $N$ containing the zero padded scaling filter coefficients for unit level,
defined via $g_{1,l} = \left(-1\right)^{l+1} \ h_{1,L-1-l}$ for $l=0, \ldots, L-1$ with
DFT $\left\lbrace G_{1,k}, k=0,\ldots, N-1 \right\rbrace$. To obtain the level $1$ wavelet 
and scaling coefficients, $\mathbf{X}$ is filtered using $h_{1}$ and $g_{1}$, and then resampled by 2
\begin{equation*}
\begin{array}{c}
W_{1,n} = \sum\limits_{l=0}^{\min(L,N)-1}\ \  h_{1,l} \ X_{2(n+1)-1-l \ {\mathrm{mod}} \ n} \\
V_{1,n} = \sum\limits_{l=0}^{\min(L,N)-1}\ \  g_{1,l} \ X_{2(n+1)-1-l \ {\mathrm{mod}} \ n} \\
\end{array}
\end{equation*}

where  $n=0,\ldots,N/2-1$ and 
$\mathbf{h}_{1}$ and $\mathbf{g}_{1}$ can be considered as high pass and low pass filters respectively.
Therefore, the wavelet coefficients $\mathbf{W}_{1}$ represents the coarse component of $X$ and
the scaling coefficients  $\mathbf{V}_{1}$ the smooth part of $X$. \\
To obtain the level $2$ wavelet $\mathbf{W}_{2}$ and scaling $\mathbf{V}_{2}$
coefficients $\mathbf{V}_{1}$ is high pass and low pass filtered as before using $h_{1}$ 
and $g_{1}$ and then subsampled by  2. Finally, at level $j$,  $\mathbf{W}_{j}$ and  $\mathbf{V}_{j}$
are given by subsampling by 2 the high pass and low pass filtering of $\mathbf{V}_{j-1}$ using $h_{1}$ 
and $g_{1}$. Here, the time series is considered to be circular and as a consequence
the first $L-2$ coefficients at each level $j$ are affected by boundary conditions and therefore
their statistical properties differ from those unaffected by circularity.
This algorithm is known as the DWT pyramid algorithm (\cite{mallatpyramid}).  \\

It is convenient
to define equivalent $\mathbf{h}_{j}$ and $\mathbf{g}_{j}$ filters with $L_{j}=(2^{j}-1)(L-1)+1$ non 
zero elements such that
\begin{equation}
\label{jlevelDWT}
\begin{array}{c}
W_{j,n} = \sum\limits_{l=0}^{\min(L_{j},N)-1}\ \  h_{j,l} \ X_{2^{j}(n+1)-1-l \ {\mathrm{mod}} \ n},\  \   n=0,\ldots,N_{j}-1 \\
V_{j,n} = \sum\limits_{l=0}^{\min(L_{j},N)-1}\ \  g_{j,l} \ X_{2^{j}(n+1)-1-l \ {\mathrm{mod}} \ n}, \  \  n=0,\ldots,N_{j}-1 \\
\end{array}
\end{equation}
and which are respectively the inverse DFT of
\begin{equation}
H_{j,k}=H_{1,2^{j-1}k \ {\mathrm{mod}} \ N \ } \ \prod_{l=0}^{j-2} G_{1,2^{l}k \ {\mathrm{mod}} \ N }, \ \ k=0, \ldots, N-1
\end{equation}
and
\begin{equation}
 G_{j,k} = \prod_{l=0}^{j-1} G_{1,2^{l}k \ {\mathrm{mod}} \ N}, \  \  k=0, \ldots, N-1
\end{equation}
This equivalent filter representation permits a clear understanding of the overall filtering applied to the
time series to obtain the wavelet and scaling coefficients at level $j$. Within this framework
it is obvious that the coarser components of $X$ are represented by the low $j$ wavelet coefficients
and the smoother ones by the large $j$ wavelet coefficients and the scaling coefficients. Indeed, the
DWT leads to a dyadic sampling in frequency so that the wavelet coefficients are obtained by band
pass filtering of $X$ in the frequency interval ${\mathcal{I}}{j}  \equiv 
\frac{1}{2^{j+1}} \leq \left|f\right| \leq  \frac{1}{2^{j}}$
for unity sampling frequency.\\

Finally, the DWT can be also represented as the decomposition of the time series into an orthonormal base
\begin{equation}
 X(t_{m}) = \sum\limits_{j=1}^{J} \ \sum\limits_{k=0}^{N/2^{j}-1} W_{j,k} \ \psi_{j,k} (t_{m}) + 
             \sum\limits_{k=0}^{N/2^{J}-1}  V_{J,k} \ \phi_{j,k} (t_{m}) 
\end{equation}
and therefore, the {\it total energy} is conserved
\begin{equation}
\left\| \mathbf{X} \right\|^{2} = \sum\limits_{j=1}^{J}  \left\|\mathbf{W}_{j}\right\|^2 + \left\|\mathbf{V}_{J}\right\|^2
\end{equation}
where $\psi_{j,k}$ and $\phi_{j,k}$ are respectively the wavelet and scaling functions.

\subsection{Wavelet bases}
\label{wavebasessection}
In this paper we concentrate on compactly supported orthogonal wavelet bases such as the Haar, Daubechies
Coifflets, Least Asymmetric (also called symmlets) and Best Localized wavelets which are extensively used in the statistical
community (see for example \cite{daubbook,percibook,vidakovicbook}). 
The Haar wavelet and scaling filters are given by
\begin{equation}
\begin{array}{l}
 h^{Haar} = \left\lbrace 1/\sqrt{2}, -1/\sqrt{2} \right\rbrace  \\
 g^{Haar} = \left\lbrace 1/\sqrt{2}, 1/\sqrt{2} \right\rbrace
\end{array}
\end{equation}
From this definition, it is obvious that the $h^{Haar}$ filter is a high-pass filter and the wavelet coefficients 
are given by the difference between adjacent data samples. In addition, the $g^{Haar}$ filter is clearly a low-pass filter 
and the scaling coefficients are given by the mean value of adjacent data samples. \\

The so called Daubechies, Coifflets, Least Asymmetric and Best Localized wavelet bases correspond each of them to
a different wavelet family. Each wavelet family is defined to fulfill a particular requirement like
for example the symmetry of the wavelet function in the case of the Least Asymmetric family. For each family the width, 
$L$, of the scaling and wavelet filters is a variable parameter and as discussed by \cite{laipaper} it is
directly related to the shape of the low-pass and high-pass equivalent filters. In particular, the larger $L$ the
closer the low and high pass filters will be to a perfect low pass and high pass filter respectively. This will be
discussed in more details in the following sections.
\subsection{The Maximal Overlap DWT}
\label{wavemodtransform}
The Maximal Overlap DWT (MODWT) also known as the non-decimated DWT or the stationary DWT, is a particularly
interesting generalization of the DWT. Actually, it allows us to keep the original sampling in time of
the time series at each level $j$ of the wavelet transform. In addition, the MODWT is invariant with
respect to time shifting while the DWT is not. These two properties will reveal important when studying
the time evolution properties of the timeline. \\

From a mathematical point of view the level $J_{0}$
MODWT of a time series $X_{t}$ of $N$ data samples is a transform consisting of $J_{0}+1$ vectors 
$\widetilde{{\mathbf{W}}}_{1}, \ldots,\widetilde{{\mathbf{W}}}_{J_{0}}$ and $\widetilde{{\mathbf{V}}}_{J_{0}}$,
all of which have dimension $N$. The vector $\widetilde{{\mathbf{W}}}_{j}$ contains the MODWT wavelet
coefficients associated with changes on scale $\tau_{j} \equiv 2^{j-1}$ while the 
$\widetilde{{\mathbf{V}}}_{J_{0}}$ contains the MODWT scaling coefficients associated with 
averages on scales  $\lambda_{J_{0}} \equiv 2^{J_{0}}$. \\

As for the DWT $\widetilde{{\mathbf{W}}}_{j}$ and $\widetilde{{\mathbf{V}}}_{j}$ can be calculated by
filtering $X_{t}$, namely,
\begin{equation}
\begin{array}{l}
\widetilde{{\mathbf{W}}}_{j} \equiv \sum_{l=0}^{L_{j}-1} \ {\widetilde{h}}_{j,l} X_{t-l \ mod \ N} \\
\widetilde{{\mathbf{V}}}_{j} \equiv \sum_{l=0}^{L_{j}-1} \ {\widetilde{g}}_{j,l} X_{t-l \ mod \ N} 
\end{array}
\label{modwtequ}
\end{equation}
$t=0, \ldots, N-1$, where $\left\lbrace {\widetilde{h}}_{j,l} \right\rbrace$  and
$\left\lbrace {\widetilde{h}}_{j,l} \right\rbrace$ are the $j$th level MODWT wavelet and scaling filters.
These filters are defined in terms of the DWT wavelet and scaling filters as
${\widetilde{h}}_{j,l} = h_{j,l}/2^{\frac{j}{2}}$ and ${\widetilde{g}}_{j,l} = g_{j,l}/2^{\frac{j}{2}}$
and have width $L_{j} \equiv (2^{j}-1)(L-1)+1$. In practice the wavelet and scaling coefficients are not
calculated from equation \ref{modwtequ} but via a pyramidal algorithm similar to the DWT
(\cite{percibook,vidakovicbook}). Note that the first $min\left\lbrace L_{j}-2,N-1 \right\rbrace$ elements
in both $\widetilde{{\mathbf{W}}}_{j}$ and $\widetilde{{\mathbf{V}}}_{j}$ are affected by circularity
and therefore are boundary coefficients. \\

Unlike the DWT, the MODWT is not an orthonormal transform because at each level $j$ the components of
both $\widetilde{{\mathbf{W}}}_{j}$ and $\widetilde{{\mathbf{V}}}_{j}$ are not independent. Nevertheless,
the MODWT is capable of producing a multi-resolution analysis of the data such that
\begin{equation}
\left\| \mathbf{X} \right\|^{2} = 
\sum\limits_{j=1}^{J}  \left\| {\widetilde{\mathbf{W}}}_{j} \right \|^2 + 
\left\| {\widetilde{\mathbf{V}}}_{J} \right\|^2
\end{equation}
Finally, the MODWT can also be expressed as a matrix operations such that 
$\widetilde{{\mathbf{W}}}_{j} = \widetilde{{\mathcal{W}}}_{j} \mathbf{X}$ and
$\widetilde{{\mathbf{V}}}_{j} = \widetilde{{\mathcal{V}}}_{j} \mathbf{X}$. The time series $\mathbf{X}$
can be recovered from 
$$
\mathbf{X} = \sum_{j=1}^{J} \widetilde{{\mathcal{W}}}_{j}^{T} \widetilde{{\mathbf{W}}}_{j}  +
               \widetilde{{\mathcal{V}}}_{J}^{T} \widetilde{{\mathbf{V}}}_{J} 
$$

\subsection{The Discrete Wavelet Packet Transform}
\label{waveptransform}
The wavelet transform as defined above is just a representation of temporal signals at different
scales $j$ which correspond to the frequency intervals $\left\lbrace \frac{1}{2^{j+1}},\frac{1}{2^{j}}\right\rbrace$ .
This representation although extremely useful and powerful, is unusual from the Fourier point of view. 
At this respect, it is interesting to define the discrete wavelet packet transform, DWPT, by generalizing the DWT, so that  
we can obtain an equipartition of the frequency domain and of the time domain similar to that given by 
a Windowed Fourier Transform (WFT). In simple words, the DWPT is obtained at each level $j$ by high pass filtering with
$\left\lbrace h_{1,l} \right\rbrace$ and low pass with $\left\lbrace g_{1,l} \right\rbrace$ the smooth 
component ${\mathbf{V}}_{j-1}$ in the DWT, but also filtering high pass and low pass the coarse
component ${\mathbf{W}}_{j-1}$. \\

The level $j$ DWPT of a $N=2^{J}$ dimensional vector $\mathbf{X}$
is an orthonormal transform yielding an $N$ dimensional vector of coefficients that can be partitioned as
$$\left\lbrace {\mathbf{W}}_{j,n}, \ \ n=0,\ldots,2^{j}-1 \right\rbrace$$
where each ${\mathbf{W}}_{j,n}$ has dimension $N_{j}=2^{J-j}$ and is nominally associated
with the frequency interval ${\mathcal{I}}_{j,n} \equiv \left\lbrace \frac{n}{2^{j+1}} \frac{n+1}{2^{j+1}}\right\rbrace$.
Note that we define ${\mathbf{W}}_{0,0}=\mathbf{X}$ following previous definitions.
Together these $2^{j}$ vectors divide the frequency interval $\left\lbrack 0,\frac{1}{2}\right\rbrack$ 
into $2^{j}$ intervals of equal width $\frac{1}{2^{j+1}}$. This leads to an equipartition both in time and
frequency of the time-frequency plane similar to that produced by the WFT without explicit time partitioning.

Defining $W_{j,n,t}$ the th element of ${\mathbf{W}}_{j,n}$ we can write
\begin{equation}
\label{dwptequ}
W_{j,n,t}=\sum_{l=0}^{L-1} u_{n,l} \ W_{j-1, \lfloor \frac{n}{2}\rfloor, 2t+1-l  
\ {\mathrm{mod}} \ N_{j-1}}, \ \ \ t=0,\ \ldots \ ,N_{j}-1
\end{equation}
where
\begin{equation}
u_{n,l} \equiv \left\lbrace
\begin{array}{l}
g_{1,l}, \mathrm{if} \ \  n \ \ \mathrm{mod} \ \ 4 =0 \ \ \mathrm{or} \ \ 3; \\
h_{1,l}, \mathrm{if} \ \ n \ \ \mathrm{mod}  \ \ 4 =1 \ \ \mathrm{or} \ \ 2; \\
\end{array}
\right.
\end{equation}
and $\lfloor \frac{n}{2}\rfloor$ is $\frac{n}{2}$ if n is even and otherwise it is $\frac{n-1}{2}$. \\

Since the DWPT is an orthonormal transform, we can use it to partition the energy in $\mathbf{X}$ via
$$\left\| \mathbf{X} \right\|^{2} = \sum\limits_{n=0}^{2^{j}-1}  \left\|\ {\mathbf{W}}_{j,n} \right\|^2$$
where $\left\|\ {\mathbf{W}}_{j,n} \right\|^2$ can be interpreted as the contribution to the energy due
to frequencies in the band  ${\mathcal{I}}_{j,n}$. As the above relationship is valid for any given level
$j$ in the interval $ 1 \leq j \leq J$, we can consider the DWPT for $ 1 \leq j \leq J$ is an overcomplete
representation of $\mathbf{X}$. Further the basis functions associated to it constitute a sort dictionary
which is commonly used for representing functions via Matching Pursuit techniques (\cite{matchingpursuitref}).

\section{Wavelet analysis of Gaussian stationary noise}
\label{statnoise}
The analysis of stationary Gaussian processes is traditionally performed in the Fourier domain because
of the decorrelation properties of the Fourier transform. Indeed, the covariance matrix, $\hat{C}_{X}$, 
of stationary Gaussian distributed noise, $X_{t}$, is diagonal in the Fourier space and 
completely described by the noise power spectrum, $P_{X}$,
\begin{equation}
\hat{C}_{X}(f,f^{\prime}) = diag \left\lbrace P_{X} (f) \right\rbrace
\end{equation}
This property makes trivial the calculation of the inverse of the covariance matrix and from the
point of view of CMB analysis, it eases considerably tasks like map making and
angular power spectrum estimation.
However, this feature can not be extrapolated to non stationary Gaussian noise. 
In this section, we show how the DWT can also decorrelate Gaussian stationary noise. Therefore,
based on the localization properties of the DWT we can reasonably expect it will also decorrelate
non stationary Gaussian noise as discussed in the following sections. 

\subsection{Correlation between wavelet coefficients.}
Using equation \ref{jlevelDWT} and assuming a stationary process $X_{t}$ with autocovariance
\begin{equation}
s_{X,\tau} \equiv E \left\lbrace X_{t},X_{t,\tau}\right\rbrace 
\end{equation}  
which depends only on $\tau$ we can write the correlation between non boundary wavelet coefficients as follows 
\begin{equation}
\begin{array}{l}
cov \left\lbrace W_{j,t}, \ W_{j^{\prime},t^{\prime}} \right\rbrace =  \\
\sum_{l=0}^{L_{j}-1} \sum_{l^{\prime}}^{L_{j^{\prime}}-1} \ h_{j,l} h_{j^{\prime},l^{\prime}} \
s_{X,2^{j}(t+1)-l-2^{j^{\prime}}(t^{\prime}+1)+l^{\prime}} \\
\end{array}
\end{equation}
which using the DFT transforms into
\begin{equation}
\begin{array}{l}
cov \left\lbrace W_{j,t}\ W_{j^{\prime},t^{\prime}} \right\rbrace = \\
\int_{-\frac{1}{2}}^{\frac{1}{2}} \ e^{i2\pi \ f \left( 
2^{j^{\prime}} \left( t^{\prime}+1\right)-2^{j} \left( t+1\right)\right)}
 H_{j}(f) \  H_{j^{\prime}}^{*}(f)
  \ S_{X} (f)  \ df \\
\end{array}
\label{wavcoeffgencorr}
\end{equation}
where $S_{X} (f)$ is the spectral density function of $X_{t}$.
Under the assumption of nominal band pass filters, $H_{j}(f)$ and $H_{j^{\prime}}(f)$ have
band-passes given by ${\mathcal{I}}_{j}$ and ${\mathcal{I}}_{j^{\prime}}$ which for $j \neq j^{\prime}$
do not overlap and thus the correlation of wavelet coefficients across scales should be negligible for a 
Gaussian stationary process. \\

Further, when $j=j^{\prime}$ and $t^{\prime}=t+\tau$ the correlation between wavelet
coefficients reads
\begin{equation}
\begin{array}{l}
C_{W} (j, \tau) = E \left\lbrace W_{j,t}\ W_{j,t+\tau} \right\rbrace = \\
\int_{-\frac{1}{2}}^{\frac{1}{2}} \  e^{i2^{j+1}\pi f \tau }
{\mathcal{H}}_{j}(f)  \ S_{X} (f)  \ df \\
\end{array}
\label{wavcoeffcorrfunct}
\end{equation}
Once again under the approximation of nominal band-pass filters 
we can argue that the above should be approximately zero for $\tau \neq 0$  when $S_{X}(f)$ is approximately
constant in the interval ${\mathcal{I}}_{j}$. In other words, at each level $j$ the wavelet coefficients
${\mathbf{W}}_{j}$ behaves as uncorrelated white noise the same way Fourier coefficients do. \\

As discussed above, the transfer function of the wavelet filter
${\mathcal{H}}_{j} = H_{j} \ H_{j}^{*}$ at each level $j$ can be well approximated by a nominal
band-pass filter taking the form
\begin{equation}
{\mathcal{H}}_{j}(f) = \left\lbrace
\begin{array}{ll}
 2^{j} & \frac{1}{2^{j+1}} \leq \left|f\right| \leq  \frac{1}{2^{j}} \\
 0     &  otherwise
\end{array}
 \right.
\end{equation}
Within a particular wavelet family, this approximation will be more accurate for large filter widths $L$, however the
number of boundary coefficients $n_{b} = L-2$ will be more important. In practice, it is always possible to find
a relative low value of $L$ for which the wavelet filter acts nearly as a perfect high pass on the data such that
the decorrelation between wavelet coefficients takes place. \\

In summary, for a stationary Gaussian process $X_{t}$ with spectral density
function $S_{X} (f)$ and under the nominal band-pass filter approximation the wavelet
coefficients at scale $j$ are uncorrelated with variance
\begin{equation}
\begin{array}{l}
  C_{j} = var \left\lbrace {\mathbf{W}}_{j} \right\rbrace = 
          \int_{\frac{-1}{2}}^{\frac{1}{2}}  {\mathcal{H}}_{j}(f) \ S_{X}(f) \ df \simeq \\
          \frac{1}{\frac{1}{2^{j}}-\frac{1}{2^{j+1}}} \int_{\frac{1}{2^{j+1}}}^
         {\frac{1}{2^{j}}} S_{X}(f)\ df \\
\end{array}
\label{wavcoeffvariance}
\end{equation}
We have ignored the scaling coefficients.
However, we can also calculate their variance from the r.m.s. of the timeline
so that
\begin{equation}
 C_{J+1}= var \left\lbrace {\mathbf{V}}_{J}\right\rbrace = 
 N \times \left\lbrace var \left\lbrace X_{t} \right\rbrace -
 \sum_{j=1}^{J} \frac{var \left\lbrace  {\mathbf{W}}_{j} \right\rbrace}{2^{j}} \right\rbrace
\label{scalcoeffvariance}
\end{equation}

\subsection{Wavelet description of $1/f$--type noise}

 \begin{figure*}[t]

 \caption{From left to right and from top to bottom, input power spectrum of simulated
$1/f$-type noise, histogram of the simulated noise wavelet coefficients and their expected values at scale $j=1$,
correlation of the wavelet coefficients and their expected values for scale $j=1$ and $j=10$ and cross correlation
of the wavelet coefficients at $j=1$ and $j=10$ with scales $j=2$ and $j=11$ respectively.}
 \label{fig:waveletdecorr}
 \end{figure*}

For most CMB data sets the noise associated to each detector timeline corresponds generally to a
stationary long memory random process also known as $1/f$--type noise. This kind of processes 
present long term correlations in time which show up in the Fourier domain as a strong
increase of power with decreasing frequency. The specific characteristics of the noise depends very much
on the detector itself as well as on the operational conditions like for example the temperature of the
focal plane bath, the current and intensity applied to the detector, the electronics, etc.
In general, the noise properties of given detector are badly known and directly derived from the
data themselves. For large data sets as would be the case for the PLANCK satellite surveyor, this
task can be extremely expensive in computation and requires fast techniques and algorithms. 
In this section we show how the wavelet transform can be used to represent and study $1/f$--type noise 
based on the previous results. The DWT has two main advantages with respect to the DFT. First, from the
computationally point of view it is faster and presents no limits in the number of samples to be used as
the DFT do; and secondly, the wavelet representation of stationary process is more compact
in terms of number of coefficients. \\

The simplest $1/f$--type noise is the white noise which is completely uncorrelated in the time
domain and presents a flat spectral density function of the form 
\begin{equation}
\label{wneq}
S^{WN} (f) = \sigma^{2}
\end{equation}
Under the hypothesis of nominal band-pass wavelet filters and using equation \ref{wavcoeffvariance} the variance 
of the white noise wavelet coefficients at scale $j$ is given by
\begin{equation}
 C^{WN}_{j} = \sigma^{2}
\end{equation}
Using equation \ref{wavcoeffcorrfunct} we can show that the white noise wavelet coefficients are fully
uncorrelated at each scale j
\begin{equation}
 C_{W}^{WN} (j,\tau) = \left\lbrace  
 \begin{array}{ll}
  C^{WN}_{j} & if \ \tau = 0 \\
  0          & otherwise 
 \end{array} 
 \right. 
\end{equation}
As the white noise wavelet coefficients are also uncorrelated between different scales 
$j$ and $j^{\prime}$ we can state that the wavelet transform of a white noise is a also 
a white noise of the same spectral density function. \\

In general the spectral density function of $1/f$--type noise is represented as
\begin{equation}
S_{X} (f) = \sigma^{2} \left( 1 + {\left( \frac{f_{knee}}{f} \right)}^{\alpha} \right)
\label{oneoverfnoisespetrum}
\end{equation}
where $f_{knee}$ is known as the knee frequency. For $f > f_{knee}$ the spectral density
function is dominated by the white noise term which is parametrized by $\sigma$. By contrast,
for $f < f_{knee}$ the $1/f$ dominates. From equation \ref{wavcoeffvariance}
the variance of the wavelet coefficients at level $j$ for $1/f$--type noise is given by
\begin{equation}
C^{1/f}_{j} = \left\lbrace
  \begin{array}{ll}
   \sigma^{2} \left[ 1 + 2^{j+1} f_{knee} \ln (2) \right] & if\ \alpha = 1 \\ 
   \sigma^{2} \left[ 1 + \frac{f_{knee}^{\alpha}}{1 - \alpha} \left( 2^{(1-\alpha)} - 1 \right) 
   2 ^{\alpha (j+1)} \right] & if\ \alpha > 1
 \end{array} 
 \right.
\end{equation}
Further using equation \ref{wavcoeffcorrfunct} the correlation between wavelet coefficients is given by
\begin{equation}
\begin{array}{l}
 C_{W}^{1/f}(j,\tau)  = C_{W}^{WN} (\tau) + \sigma^{2} \ 2^{(j+1)} \ f_{knee}^{\alpha} \ \ \ldots \\
 \ldots \ \int_{2^{-j-1}}^{2^{-j}} \cos \left(2^{j+1} \pi f \tau \right) \ f^{-\alpha} \ df
\end{array}
\end{equation}
which is largely dominated by the white noise term so that the wavelet coefficients are quasi-uncorrelated.
This equation is only valid for nominal band-pass filters. However, from a practical point of view increasing
the number of non-zero coefficients in the wavelet filter will flatten the $1/f$ term in the integral leading
to a desired level of decorrelation between wavelet coefficients. \\

To test the above statements we have simulated a TOD of $1/f$-type noise with $2^{15}$ samples. 
Figure~\ref{fig:waveletdecorr} shows from left to right and from top to bottom, the input power spectrum of simulated $1/f$-type noise, the histogram of the simulated noise wavelet coefficients at scale $j=1$,
the correlation of the wavelet coefficients and their expected values for scale $j=1$ and $j=10$ and the cross correlation
of the wavelet coefficients and their expected values at $j=1$ and $j=10$ with scales $j=2$ and $j=11$ respectively.
We observe that the wavelet coefficients show a Gaussian structure. Further, they are quasi-decorrelated
as expected within the same scale and totally uncorrelated between scales. 

\subsection{Simulation of Gaussian stationary noise using wavelets.}
\label{wavsimugaussnoise}

 \begin{figure*}[t]

 \caption{Left panel: simulated $1/f$-type noise using a wavelet algorithm. Right panel: power
spectrum of the wavelet simulated $1/f$-type noise compared to the input power spectrum in red.}
 \label{fig:waveletsim1overf}
 \end{figure*}

From the above discussion we have conclude that under the nominal band-pass filter approximation
the DWT decorrelates Gaussian stationary processes. Thus it is possible to simulate Gaussian stationary 
processes via its DWT (\cite{wavestrapping,vidakovicbook}), the same way it is performed using the DFT. 
Indeed we can easily simulate stationary and Gaussian time series $X_{t}$ of length $N=2^{J}$ and spectral density function
$S_{X}(f)$ performing the following steps 
\begin{enumerate}
\item{Using a random-number generator, generate a vector ${\mathbf{Z}}_{M}$ containing $M=4N$ Gaussian white
noise deviates with zero mean and unit variance}
\item{Integrate numerically equation \ref{wavcoeffvariance} to calculate the approximate wavelet coefficient variances 
$C_{j}, \ \ j=1, \  \ldots \ , J+2$ and use equation \ref{scalcoeffvariance} to calculate the scaling coefficient 
variances.}
\item{Multiply the first $M/2$ elements of ${\mathbf{Z}}_{M}$ by $\sqrt{C_{1}}$; the next $M/4$ values by $\sqrt{C_{2}}$
and so forth, until to the final four elements, which need to be multiply by $\sqrt{C_{J+1}}$,
$\sqrt{C_{J+1}}$, $\sqrt{C_{J+2}}$, $\sqrt{C_{J+3}}$ respectively.}
\item{Compute the inverse DWT of the modified  ${\mathbf{Z}}_{M}$ vector and then randomly choose $M$ samples from it.}
\end{enumerate}  

The above recipe can be easily understood in terms of the wavelet covariance matrix of a Gaussian stationary time series
with $N=2^{J}$ samples which due to the decorrelation property is given by 
\begin{equation}
\begin{array}{l}
\Sigma_{{\mathbf{W}}} = 
 diag \ \lbrace \ \underbrace{C_{1},  \ldots, C_{1}}_{\frac{N}{2} \  of \ these},
 \underbrace{C_{2},  \ldots, C_{2}}_{\frac{N}{4}\ of \ these} , \ \ldots \ , \\ 
 \underbrace{C_{j},  \ldots, 
C_{j}}_{\frac{N}{2^{j}} of these}, 
 \ldots , \underbrace{C_{J-1}, C_{J-1}}_{2 \  of \ these}, C_{J}, C_{J+1}
\ \rbrace \\
\end{array}
\label{wavcoeffcovariance}
\end{equation}

The left panel of figure~\ref{fig:waveletsim1overf} shows a simulation of $1/f$-type noise obtained
from the algorithm described above. In the right panel, we represent the power spectrum of the
simulated data which is in agreement with the input power spectrum plotted in red.

\section{Locally stationary Gaussian noise.}
\label{localstatnoise}

 \begin{figure*}[t]
 \caption{From left to right, locally stationary Gaussian noise obtained from two
white noises of variances 1 and 25, and its wavelet transform with wavelet coefficients
ordered from left to right for decreasing scale.}
 \label{fig:localstatwhitenoise}
 \end{figure*}

Locally stationary processes appear in many physical systems in which the mechanisms that produce
random fluctuations change slowly in time. Over given time intervals, such processes can
be approximated by a stationary one. This is the case, for example, of the variations observed in 
the noise total power of some CMB detectors due to fluctuations in the focal plane temperature
as discussed in subsection~\ref{archtimefreq}.
From the Fourier point of view we can imagine the variation of the data power spectrum with 
time. Actually, we could also define locally-stationary Gaussian processes as those for which the
time-frequency plane is divided into time intervals corresponding each of them to a 
stationary process and which are uncorrelated between them.  Notice that the above definitions
of locally-stationary processes are very general and in the limit of infinitesimal time intervals
they lead to non-stationarity in a more general manner.
\subsection{Fourier approximation to locally-stationary process}
For a zero-mean random process ${\mathbf{X}}_{t}$ we can define its time evolving autocovariance as
\begin{equation}
s_{{\mathbf{X}}}(t,t^{\prime})= E \left\lbrace {\mathbf{X}}(t)\ , \ {\mathbf{X}}(t^{\prime}) \right\rbrace
\end{equation}
which can also be expressed in terms of the distance between $t$ and $t^{\prime}$ and the mid-point 
between them
\begin{equation}
s_{{\mathbf{X}}}(t,t^{\prime})= C (\frac{t+t^{\prime}}{2},t-t^{\prime})
\end{equation}
Note that under this definition the covariance of a stationary process reads
\begin{equation}
C (\frac{t+t^{\prime}}{2},t-t^{\prime}) = C (t-t^{\prime}) = C (\tau)
\end{equation}

For a locally stationary process, we expect that in the neighborhood of any $x \in \mathcal{R}$,
there exists an interval of size $l(x)$ where the process can be approximated by a stationary one.
Therefore for any $\frac{t+t^{\prime}}{2} \in \left\lbrack x-\frac{l(x)}{2}, 
x+\frac{l(x)}{2}\right\rbrack$
\begin{equation}
C (\frac{t+t^{\prime}}{2},t-t^{\prime}) \approx C (x,t-t^{\prime})
\end{equation}
Further, if $t$ and $t^{\prime}$ are far apart and do not belong to the same interval of stationarity
then we expect that $X(t)$ and  $X(t\prima)$ are uncorrelated. This means for any $u \in  \left\lbrack x-\frac{l(x)}{2}, 
x+\frac{l(x)}{2}\right\rbrack$, if $\left| v \right| > \frac{l(x)}{2}$ then
\begin{equation}
 C (u,v) = E\left\lbrace X(u+\frac{v}{2}), X(u-\frac{v}{2})\right\rbrace \approx 0
\end{equation}

In the time frequency-plane we can define a time-varying spectrum $\Lambda(t,w)$ which would be given 
by the Fourier transform of the time-varying covariance $C (t+\frac{v}{2},t-\frac{v}{2})$ with 
respect to $v$. \\
For stationary processes $\Lambda(t,w)$ does not depend on time and corresponds to
the standard power spectrum. For locally-stationary processes (\cite{adaptivecovest}) the spectrum
varies with time but it is constant in time within the stationarity intervals defined above.  
This means that the same way the Fourier transform decorrelates Gaussian stationary processes,
the windowed Fourier transform decorrelates locally stationary processes. In other words, the covariance
matrix of locally-stationary Gaussian noise is diagonal in an orthogonal base given by performing a
windowed Fourier transforms in each stationary interval (\cite{adaptivecovest}). This base divides the
time-frequency plane in rectangles of the form 
$$ \left\lbrack x-\frac{l(x)}{2},x+\frac{l(x)}{2}\right\rbrack \ \times \  
\left\lbrack f_{k}-\frac{1}{2 \ l(x)},f_{k}+\frac{1}{2 \ l(x)}\right\rbrack $$
where $f_{k} = \frac{k+ \frac{1}{2}}{l(x)}, \ \ k=1,2,\ldots, \frac{f^{samp}}{2 \ l(x)}$ and $f^{samp}$
is the sampling frequency. Note that the DWPT produces a similar decomposition of the time-frequency
plane and can also be used to obtain a diagonal base for the covariance matrix of stationary
Gaussian process $\mathbf{X}_{t}$. For this purpose, best basis algorithms combining the DWPT
at all available scales $j$ have been developed (\cite{entbestbasis,percibook,vidakovicbook}). 
We will not account for them in this paper. \\

In the previous section we have shown that the DWT decorrelates Gaussian stationary processes under the
approximation of nominal band-pass wavelet and scaling filters. Therefore, the DWT will also decorrelate
a locally-stationary Gaussian process which can be considered as series of attached stationary Gaussian processes 
uncorrelated between them. The same way we have defined above a time varying power spectrum we can also consider
the evolution in time of the variance of the wavelet coefficients at each level $j$. As wavelet coefficients are 
localized in time, this evolution can be naturally computed from the DWT of the data. Thus, for Gaussian
locally-stationary processes we expect the variance of the wavelet coefficients to change from one stationary
interval to another and to be constant within a given interval. This statement can be fully understood 
considering equation \ref{wavcoeffvariance} and accounting for the time evolution of the data power spectrum.
To clarify this issue we can write, from equation \ref{wavcoeffcovariance}, the covariance matrix of the 
wavelet coefficients of a Gaussian locally-stationary process composed of two stationary time intervals 
\begin{equation}
\begin{array}{l}
\Sigma_{{\mathbf{W}}} =
 diag \ \lbrace  
 \ \underbrace{C^{1}_{1},C^{1}_{1}}_{\frac{N}{4} \ of \ these}
 ,  \ldots, \underbrace{C^{2}_{1},C^{2}_{1}}_{\frac{N}{4} \  of \ these},
 \underbrace{C^{1}_{2},C^{1}_{2}}_{\frac{N}{8} \ of \ these},  \ldots, \\
 \underbrace{C^{2}_{2},C^{2}_{2}}_{\frac{N}{8} \ of \ these} , \ldots,  

 \underbrace{C^{1}_{j},C^{1}_{j}}_{\frac{N}{2^{j+1}} \ of \ these}, \ldots, \\
 \underbrace{C^{2}_{j},C^{2}_{j}}_{\frac{N}{2^{j+1}} \ of \ these}, 
 \ldots , \underbrace{C^{1}_{J-1}}_{1 \ of \ these}, \underbrace{C^{2}_{J-1}}_{1 \ of \ these}, 
  C_{J}, C_{J+1} \ 
\rbrace
\end{array}
\label{wavcoeffvarlocallystat}
\end{equation}
where the indexes $1$ and $2$ refer to each of the stationary time intervals respectively. In summary, the
DWT decorrelates Gaussian locally-stationary processes,  and  naturally permits the identification of
the distinct intervals of stationarity in the data using the time evolution of the wavelet coefficient variance. \\

For illustration, we show in the left plot of figure~\ref{fig:localstatwhitenoise} an example of a locally stationary noise given by the superposition in time of two white noises of variances 1 and 25. Its wavelet transform is traced in the right plot with the wavelet coefficients ordered from left to right for decreasing scale. We can clearly observe for each scale two distinct parts corresponding to the two white noises.
\subsection{Time modulated-stationary processes}
\label{sec:timemodulatedstationaryprocesses}
Often, as discussed in the introduction to this section, the noise total power of CMB instrument detectors
changes with time due to slow fluctuations on the temperature of the focal plane bath or
drifts in the background power. In general, this leads
to locally-stationary Gaussian noise for which the power spectrum varies along the observation period
but in the same way for all frequency components. Indeed, the total noise power increases and/or decreases with time 
but the shape of the power spectrum remains the same all over the observation period.
These processes can be well approximated by a stationary Gaussian process modulated on time
(\cite{timemodnoisepaper}) which is defined by
\begin{equation} 
X_{t} = \sigma(t) \  \times \ Y_{t}
\label{timemodproequ}
\end{equation}  
where $Y_{t}$ is a Gaussian stationary process and $\sigma(t)$ is a function of time. From a more physical point
of view, $\sigma(t)$ represents the change on the noise total power. For convenience we will consider in the
following that the variance function $\sigma^{2}(t)$ is defined on $(0,1)$ by assuming $Y_{t}$ has as power
spectrum the time-constant power spectrum of $X_{t}$.
Finally, notice that time modulated processes are just a particular case of locally-stationary processes. \\

From a practical point of view, the above model is quite handy because only $\sigma(t)$ has to be estimated
from the data. Simple although biased estimates of  $\sigma(t)$ can be obtained by comparing the non-time dependent power
spectrum of the time series $P_{X}(w)$ to power spectra, $P_{x}(w,t_{k})$ calculated at different but connected 
time intervals along the observation period
\begin{equation}
{\widetilde{\sigma}}^{2} (t_{k}) = \frac{\sum_{w} P_{x}(w,t_{k})}{\sum_{w} P_{X}(w)}
\end{equation}
This estimate leads to the approximation of $\sigma(t)$ by a step-function so that the time intervals
have to be particularly chosen to represent as best as possible the shape of  $\sigma(t)$.
By contrast, using the MODWT it is possible to easily obtain unbiased estimates 
of $\sigma(t)$ (\cite{timemodnoisepaper}) from 
\begin{equation}
{\widetilde{\sigma}}^{2} (t) = \sum_{j=1}^{J} \ 2^{-j} \left|{\widetilde{W}}_{j,t}\right|^{2}
\label{sigmamodwtestequ}
\end{equation}
The extra factor $2^{-j}$ can be easily understood from the normalization factor introduced in the
definition of the MODWT wavelet and scaling filters when deduced from the DWT ones. Few extra comments on equation
\ref{sigmamodwtestequ} are needed. First, $\sigma(t)$ is not defined on $(0,1)$ and has to be
renormalized to be compared to the Fourier estimate. Second, the MODWT estimate is not consistent and
has to be smeared out to reduce its variance. Third, we could also obtain an unbiased estimator from
the DWT in a similar manner but its construction is more complex due to the time undersampling performed
by the DWT as increasing levels. \\

In summary, the time-modulated stationary model for random processes can be of great interest for dealing with
Gaussian locally-stationary CMB time series which present time evolution in the noise total power but not
in their power spectrum shape. Indeed, once estimates for the $\sigma(t)$ function are available the simulation
of such processes is reduced to the simulation of a stationary Gaussian process from the power
spectrum of $X_{t}$ which is then multiplied by that function.
Further, the covariance matrix of such processes can be well approximated in the wavelet space as shown in
the following subsection. 
\subsection{Time modulated-stationary wavelet processes}
Based on the definition of time modulated-stationary processes and on the
properties of the wavelet coefficients of stationary and locally stationary processes we can also define
time modulated-stationary wavelet processes as those for which the variance of the wavelets coefficients
at level $j$ is of the form
\begin{equation}
var\left\lbrace W_{j,t} \right\rbrace = \sigma^{2}_{j}(t) \times C_{j} 
\label{timemodstatwavproequ}
\end{equation}  
where $C_{j}$ is a constant and $\sigma_{j}(t)$ is a function of time. Notice that this definition accounts
also for time modulated-stationary processes for which the different  $\sigma_{j}(t)$ functions are a
consequence of the undersampling performed in the DWT. \\

As in the above definition we have supposed the wavelet coefficients are uncorrelated their covariance 
matrix is therefore diagonal of the form
\begin{equation}
\begin{array}{l}
\Sigma_{{\mathbf{W}}} =
 diag \ \lbrace  
 \ \underbrace{\sigma_{1}(t^{1}_{1}) \ C_{1},\sigma_{1}(t^{1}_{2}) \ C_{1}}_{\frac{N}{2} \ of \ these}, \ldots,  \\
 \underbrace{\sigma_{2}(t^{2}_{1}) \ C_{2},\sigma_{2}(t^{2}_{2}) \ C_{2}}_{\frac{N}{4} \ of \ these},  \ldots, 

 \underbrace{\sigma_{j}(t^{j}_{1}) \ C_{j},\sigma_{j}(t^{j}_{2}) \ C_{j}}_{\frac{N}{2^{j}} \ of \ these}, \ldots, \\
  \underbrace{\sigma_{J-1}(t^{J-1}_{1})C_{J-1},\sigma_{J-1}(t^{J-1}_{2})C_{J-1}}_{2 \ of \ these}, 
  C_{J}, C_{J+1} \ 
\rbrace \\
\end{array}
\label{wavcoeffvartimemodstatwavpro}
\end{equation}
where $t^{j}_{k}$ represents the $k$th time sample of the level $j$ DWT.
\subsection{Statistical analysis of CMB data sets with locally stationary noise.}
\label{sec:analysisofCMBlocallystationarynoise}
For most of the
time domain processing of CMB data sets like for example subtraction of systematics 
and optimal map making we have to deal with the inversion of the noise covariance matrix.
In general, we consider the noise Gaussian and stationary or piece-wise stationary,
so that the noise correlation matrix is diagonal in Fourier space. Therefore, this can
be trivially inverted for each piece of data. 
Above, we have shown that the covariance matrix of correlated Gaussian
and non stationary time modulated noise (in particular locally stationary) is 
diagonal in the wavelet space and can be simply described
by a set of coefficients $C_{j}$, where $j$ is the wavelet scale index, modulated
by a time dependent function $\sigma_{j}(t)$. Therefore, the inversion of this
matrix is also trivial in wavelet space. \\

If we consider Maximum Likelihood algorithms like for example optimal map making,
the resolution of the system involves terms of the form $N^{-1}d$ where $N$ is the noise 
correlation matrix and $d$ a data vector. In Fourier space we perform this operation by first
taking the Fourier transform of $d$, then dividing it by the diagonal terms of the Fourier
representation of $N$ and finally transforming back to real space. Equivalently, in the
wavelet case we can write for the stationary case 
  \begin{equation}
   N^{-1}d = W^{T}\left(\left\lbrace Wd \right\rbrace_{j,k}/C_{j} \right)
  \end{equation}
where $W$ represents the direct wavelet transform, $W^{T}$ the inverse wavelet transform
and $C_{j}$ are the variances of the wavelet coefficients of the noise wavelet transform.
In the same way for the non stationary time modulated noise we can write 
 \begin{equation}
      N^{-1}d = W^{T}\left(\left\lbrace W\left(d/\sigma(t)\right)\right\rbrace_{j,k}/(C_{j})\right)
  \end{equation}
where $\sigma(t)$ is the time modulation function. \\

In general the noise covariance matrix is not known and has to be estimated directly from the
data. For the stationary case, this is simple as we only need to compute the variance
of the wavelet coefficients for each of the different scales. For non stationary time
modulated noise we can have a very good approximation using Eq.
\ref{timemodstatwavproequ} by first computing the variance of
the wavelet coefficients assuming stationary noise and then computing
the time modulation function from the global evolution of the
wavelet coefficients. \\

The top panel of figure, \ref{funct_sigmat_simu} shows in black the global
time variation of the noise of one of the Archeops bolometers which decreases
with time (a complete description of the Archeos data is presented in
Section~\ref{cmbdataanalysis}). In red we overplot
the time modulation function, $\sigma(t)$ as determined from the 
data themselves following
the above technique. Finally, the bottom panel of the figure shows the 
global time variation of a simulation of the Archeops noise. This have
been obtained by multiplying by $\sigma(t)$
a realization of a stationary Gaussian noise produced
as described in section \ref{wavsimugaussnoise}. In red we overplot the $\sigma(t)$
estimated for the simulation which behaves as the one estimated from the data.
Notice that this kind of non stationary simulations of the noise in the Archeops data
were used when computing the Archeops CMB angular power spectrum in (\cite{benoit_cl}). 
 \begin{figure}[t]
 \caption{{\it Top}:  Reconstructed time modulation function $\sigma(t)$ of the Archeops noise in bolometer 217K04.
 {\it Bottom:} Time modulation function $\sigma(t)$ for a non-stationary simulation 
of the Archeops noise in bolometer 217K04.}
 \label{funct_sigmat_simu}
 \end{figure}

\section{Wavelet destriping}
\label{sec:wavdestriping}
Data from CMB experiments are commonly limited by intrinsic $1/f$-type noise in the
detectors. For experiments with a circular scanning strategy, like Archeops, WMAP and
Planck, the $1/f$-type noise shows up like stripes in the sky maps. The techniques used
to remove those stripes are in general called destriping algorithms 
(\cite{efstathioudest,keihanendest, mainodest,poutanendest, revenudest,sbarradest}). 
These are both based on the statistical properties of $1/f$-type noise and on the fact that
the noise, by contrast to the sky signal, is not coherently projected in the maps. 

\subsection{Derivation of a wavelet destriping algorithm}

In the previous sections we have shown that the wavelet transform nearly diagonalizes 
the covariance matrix  of $1/f$-type noise. This allows us to design a destriping algorithm 
based on the wavelet transform.

We assume that the time domain data for a typical CMB experiment can be written as
\begin{equation}
d_t = s_t + n^c_t +n^w_t \label{eq1}
\end{equation}
where $s_{t}$ is the sky signal, and $n^w$ and $n^c$ are respectively a white and a
correlated (color) noise component. 
We express the projection of the time domain signal, $s_t$ into a sky map, $m_p$, 
via a pointing matrix $P_{pt}$ such that
\begin{equation}
m_p = P_{pt} s_t
\label{projecting}
\end{equation}
Therefore we can rewrite equation (\ref{eq1}) as  
\begin{equation}
d_t  = P_{tp} m_p  + n^c_t +n^w_t \label{eq2}
\end{equation} 
The main purpose of a destriping algorithm is just to remove or reduce significantly the contribution from
the correlated noise. This is obtained by maximizing the likelihood function, ${\cal L }$, over the noise and
the sky signal.

We can characterize the correlated noise $n^c$ using its wavelet transform 
\begin{equation}
\bvec{W}_{nc} = {\cal W} n^c_t
\end{equation}
for which the wavelet coefficients are Gaussian distributed and quasi-uncorrelated.
We can then write the likelihood function as follows 
\begin{eqnarray}
{\cal L } &=& P(\bvec{d}|\bvec{m})= P(\bvec{W}_w|\bvec{m})P(\bvec{W}_{nc}|\bvec{m}) \\
&\sim& exp\left[-\frac{1}{2} \left(
\bvec{W}_w^T {\tilde C}_w^{-1} \bvec{W}_w  + 
\bvec{W}_{nc}^T {\tilde C}_{nc}^{-1} \bvec{W}_{nc}   \right)  \right] \label{eq3}
\end{eqnarray}  
where ${\tilde C}_w $ and ${\tilde C}_{nc}$ are the covariance matrices in wavelet space of the correlated and white noise
respectively 
\begin{eqnarray}
{\tilde C}_w  &=& < \bvec{W}_w \bvec{W}_w^T> \\
{\tilde C}_{nc} &=& <\bvec{W}_{nc} \bvec{W}_{nc}^T >
\end{eqnarray} 
Notice that for our assumptions both matrices are diagonal.
 
By taking the log of the likelihood function (\ref{eq3}) we obtain
\begin{eqnarray} 
\chi^{2} &=&  \bvec{W}_w^T {{\tilde C}_w}^{-1} \bvec{W}_w 
+{\bvec{W}_{nc}}^{T}{{\tilde C}_{nc}}^{-1}{\bvec{W}_{nc}} \\
 &=& (\bvec{d}-{\cal W}^T \bvec{W}_{nc}- {\cal P}\bvec{m} )^{T}{C_w}^{-1} \ldots \\
& & \ldots (\bvec{d}-{\cal W}^T \bvec{W}_{nc} - {\cal P}\bvec{m}) +
 \bvec{W}_{nc}^{T}{{\tilde C}_{nc}}^{-1}{\bvec{W}_{nc}} \label{eq4}
\end{eqnarray}
whose minimization with respect to $\bvec{m}$ and  $\bvec{W}_{nc}$ leads to the following system of equations 
\begin{eqnarray}
\bvec{m} & = & ({\cal P} {\cal P})^{-1}{\cal P}^T (\bvec{d} - {\cal W}^T {\bvec W}_{nc}) \\
0 & = & - {{\tilde C}_w}^{-1}{\cal W} ( I-{\cal P} ({\cal P} {\cal P})^{-1}{\cal P}^T)(\bvec{d}-{\cal W}^T 
{\bvec{W}}_{nc})+ \ldots \\
&& \ldots {{\tilde C}_{nc}}^{-1}{{\bvec{W}}_{nc}}
\end{eqnarray}
which can be combined into 
\begin{equation}
\begin{array}{l} 
({\cal W}( I-{\cal P} ({\cal P} {\cal P})^{-1}{\cal P}^T) {\cal W}^T + {\tilde C}_w 
{{\tilde C}_{nc}}^{-1}){\bvec W}_{nc}  =  \\
 {\cal W}( I-{\cal P} ({\cal P} {\cal P})^{-1}{\cal P}^T) \bvec{d}   \\
\end{array}
\label{eq5}
\end{equation}

The solution to this system of equations can be found using iterative methods
and in particular we use a conjugate gradient method (\cite{numrec}).

\subsection{Practical implementation} 
For $1/f$-type noise the variance of the wavelet coefficients at level $j$ is given by 
\begin{equation} 
C_j =  \: {\sigma}^2 \:2^{j \alpha} + {\sigma_w}^2 \label{eq6}
\end{equation}
where ${\sigma_w}^2$ is the white noise variance and ${\sigma}^2 \:2^{j \alpha}$ is 
the variance of the correlated part with $\alpha$ being the $1/f$ exponent
as described in equation \ref{oneoverfnoisespetrum}.

When the noise dominates the sky signal, we can extract the noise 
parameters 
${\sigma_w}$, ${\sigma}$ and  $\alpha$ directly from data 
itself via Monte Carlo Markov Chain methods (\cite{Wavelet:01}).   

When the signal contribution can not be neglected, the data wavelet variance 
is given by 
\begin{equation} 
{\sigma_{b \,,j}}^2 \:=   {\sigma_{w\,,j}}^2 + \sigma_{n_c\,,j}^2 +
 {\sigma_{sky \,, j}}^2  \label{eq7}
\end{equation}

${\sigma_{b \,,j}}$ and ${\sigma_{w\,,j}}$ can be iteratively evaluated
from data using a template of the sky signal which is improved at each
iteration. From the wavelet transforms of the template and the data we obtain
\begin{eqnarray} 
{\sigma_{n_c, j}}^2 ={{\sigma}_{b\,,j}}^2 - {{\sigma}_{sky\,,j}}^2 
- {\sigma_{w \,,j }}^2 \label{eq8}
\end{eqnarray}

\subsection{Application to simulated data}
\begin{figure}[t]
\caption[Baseline reconstruction on simulated Archeops data using a wavelet based
destriping algorithm]{Baseline reconstruction on simulated Archeops data using a wavelet based
destriping algorithm. In black we trace the simulated Archeops TOD. We overplot
in blue, green and red the reconstructed baseline for the first, second and third step of
the algorithm respectively.}
\label{fig:simudesttl}
\end{figure}

The wavelet destriping was applied to simulated Archeops TOD at $545$~GHz
with $3 \times 2^{21}$ samples.
We considered two main components in data, Galactic dust emission and $1/f$-type noise.
For the former we used a template of the Galactic dust emission scaled down to
$545$~GHz provided by \cite{fds}. The $1/f$-type noise properties were deduced
from the Archeops data. 

We proceeded in three main steps. First of all, we estimated the lowest frequency components
of the data from its wavelet decomposition and subtract it. Secondly, we improved this estimation
by solving equation~\ref{eq5} iteratively starting with largest scales and adding smaller and
smaller scales progressively. Finally, we computed an approximation to the signal, $s_t$, by thresholding
the previous destriped map and deprojecting into the time domain. We then, step two was applied to
the residuals, $d_t-s_t$. 

\begin{figure}[t]
\caption{Top panel: Destriped map of the simulated Archeops data at $545$~GHz using the wavelet
destriping algorithm. Bottom panel: Residual stripes on the destriped map above.}
\label{fig:simudestmap}
\end{figure}

Figure~\ref{fig:simudesttl} shows the simulated Archeops TOD at $545$~GHz. In blue,
green and red we overplot the reconstructed noise low frequency components for step one, two and
three respectively. We observe that we improve significantly the estimated baseline in step three
reducing the stripes in the maps as shown in figure~\ref{fig:simudestmap} where
we represent the destriped map on the top panel and the residual stripes on the bottom panel.

\section{Time-frequency visualization and analysis of the Archeops TOD.}
\label{cmbdataanalysis}
In the previous sections we concentrated on the wavelet description of random Gaussian processes.
We studied the decorrelation properties of the DWT and their application to the statistical
analysis of CMB data sets and in particular to the destriping
of CMB maps. From the point of view of CMB data analysis this is of great interest 
but it is not all we can obtain from the wavelet transform. Actually, the most important
property of wavelets is their simultaneous time and frequency localization. 
At this respect, wavelet analysis
and in particular the DWPT is a fundamental tool for data visualization and characterization.
CMB time series $d_{t}$ are in general a linear combination of signal, both galactic and 
cosmological in origin,
of systematics effects like atmospheric contamination, and/or parasitic noises, and/or electromagnetic contamination, and/or glitches, and of random Gaussian noise
\begin{equation}
\begin{array}{l}
d_{t} = \alpha^{cos} s^{cos}_{t} + \alpha^{gal} s^{gal}_{t} + 
\alpha^{atm} sys^{atm}_{t}+ \\ 
\alpha^{pn} sys^{pn}_{t} + \alpha^{ec} sys^{ec}_{t}+ \ldots + n_{t} \\
\end{array}
\end{equation} 
These components are different in nature and they present different time and frequency properties. For example, the
cosmological and galactic signals only depend on the position of the sky observed which is fully given by the instrument
pointing. By contrast, the atmospheric signal depends both on the sky position and on the 
time evolution of the atmospheric conditions. Further, other systematic effects vary mainly with time and in general are related
to physical phenomena in the detector surroundings. Finally, the noise component is intrinsic to each detector
and it is often of $1/f$-type, Gaussian and locally-stationary. As the cosmological signal, the CMB emission, is in the
time domain much weaker than all other components, these have to be identified, characterized and removed with high
degree of precision before the extraction of the CMB anisotropies power spectrum. 

\subsection{Time-frequency analysis of the Archeops data}
\label{archtimefreq}
\begin{figure}[t]
\caption{{\it Top}: Time-frequency analysis of the TOD of the 217K04 Archeops bolometer using
the DWPT. {\it Bottom:} Time-frequency analysis of the expected Galactic emission in the TOD of the 217K04 Archeops bolometer using the DWPT. See text for details.}
\label{timefre_gal}
\end{figure}

\begin{figure}[t]
\caption{{\it Top}: Time-frequency analysis of the TOD of the 217T04 Archeops bolometer using
the DWPT. {\it Bottom:} Time-frequency analysis of the expected Galactic emission in the TOD of the 217T04 Archeops bolometer using the DWPT. See text for details.}
\label{timefre_gal_badbolo}
\end{figure}

In the following we present the most relevant issues in the wavelet analysis of the Archeops data set.
A more detailed description of the Archeops data and its processing can be found in (\cite{arch_pipe}). 
To simplify further discussions we will assume only four main components in the Archeops timelines: CMB emission, 
galactic and atmospheric emissions, non identified systematics and Gaussian noise. 

The top panel of figure \ref{timefre_gal} shows the DWPT
time-frequency representation of a typical Archeops detector signal. We observe mainly two components which dominate
at high and low frequencies respectively. The low frequency signal which varies significantly with time is  mainly
given by galactic emission as shown by the bottom panel of the same figure where we represent the expected
galactic emission for that detector. The circular scanning strategy of Archeops is such that during most of the
observation time we cross the Galactic plane perpendicularly and therefore the galactic emission shows
up as a spike in time. By contrast in the last hours of flight, the scanning direction is
collinear to the Galactic plane and then, the galactic emission becomes broader in the time direction although
we can still observe few frequency spikes related to intense and compact regions of the Galactic plane. 
The nearly time-fixed
structures observed at low frequency although they look very much as $1/f$ noise, are due to systematics 
mainly dominated by atmospheric emission.

The high frequency signal corresponds to the detector white noise whose power is artificially raised up with
increasing frequency when deconvolving from the bolometer time constant. We observe that the noise smoothly decreases
with time showing a clear non-stationary behavior. Indeed, as time went on during the flight the bolometer temperature
decreased therefore reducing the bolometer intrinsic noise. A detailed wavelet description and modeling of the
non-stationarity of the Archeops noise is presented in the following subsection.\\

The top panel of figure \ref{timefre_gal_badbolo} shows the DWPT of the worst Archeops bolometer (217T04)
both in terms
of noise and systematics. As above we plot in the bottom panel the DWPT of the expected Galactic signal
for this bolometer. We observe that the DWPT of 217T04 presents as above at low frequencies
the Galactic and atmospheric components and at high frequency a time varying noise. However
it is dominated by a highly time varying signal
in the frequency range 25 to 35 Hz. This signal is not observed in the data presented above and can
be clearly identified as the residuals from unknown systematic effects. We can therefore, just
by visual inspection, characterize the quality of the Archeops bolometers data and identify
which of them are badly affected by systematics. We have used this technique to identify the best, in terms of
sensitivity and low level of systematics, Archeops bolometers which were used to construct the
Archeops sky maps.

\subsection{Modeling of the non-stationarity of the Archeops noise}
\begin{figure}[t]
\caption{{\it Top}: Average power spectrum of the TOD of the 217K04 Archeops bolometer
as a function of frequency computed from its DWPT {\it Bottom:} Time evolution of the
power spectrum of the TOD of the 217K04 Archeops bolometer computed from its DWPT.}
\label{fig:func_sigmat}
\end{figure}
As discussed in the previous section, the Archeops cryostat radically increased temperature when taking off
and then cooled down slowly to achieve a nominal temperature of 95 mK for the last ten hours of flight.
This produced a fair decrease of the noise power with time which is the cause of the non stationarity of
the Archeops bolometer noise. To account for this non stationarity we modeled Archeops noise as a
time modulated-stationary wavelet process as it can be considered as locally (piece-wise) stationary noise with slowly
time varying power but for the first two hours of flight.
Following subsections \ref{sec:timemodulatedstationaryprocesses} and \ref{sec:analysisofCMBlocallystationarynoise}
we have computed the $\sigma(t)$ function for each of the Archeops bolometers. The top panel 
figure~\ref{fig:func_sigmat} shows mean wavelet power as a function of frequency for the
the Archeops bolometer 217K04. The bottom panel shows the reconstructed $\sigma(t)$ function for the same bolometer.
From these two quantities we can produce non-stationary simulations of the Archeops noise as shown in 
figure~\ref{funct_sigmat_simu}. These non stationary simulations of the Archeops bolometer noise
were used to determine the angular power spectrum of the noise in the Archeops maps which were needed for 
computing the Archeops CMB angular power spectrum in (\cite{benoit_cl}) via a MASTER-like algorithm. 

\subsection{Wavelet destriping of the Archeops data}
\begin{figure}[t]
\caption{{\it Top}: Wavelet destriped map of the Archeops 545~GHz after one iteration of the
algorithm described in section~\ref{sec:wavdestriping}  {\it Bottom:} Wavelet destriped map of 
the Archeops 545~GHz after two iteration of the algorithm.}
\label{fig:mapwavdestarch}
\end{figure}

\begin{figure}[t]
\caption{Performance of the wavelet filtering in the 353~GHz Archeops data. In orange we trace
the reconstructed data baseline using a fit to a base of Gabor atoms. In blue we represent
the reconstructed baseline using the wavelet detrending algorithm.}
\label{fig:wavdetrendingtimeline}
\end{figure}

We have applied the wavelet destriping algorithm presented 
in section~\ref{sec:wavdestriping} to the Archeops data at 545~GHz. When working with
the simulated Archeops data we have only considered two main components, Galaxy emission
and $1/f$-type noise. As shown above, for the Archeops data we also need to consider 
the atmospheric emission at low frequencies which mimic a $1/f$-type but it is not Gaussian
distributed. To overcome this problem, we have first estimated and removed a very low frequency
baseline in the data. After that,
we have applied two iterations of the wavelet destriping algorithm as described in section~\ref{sec:wavdestriping}.
The results are presented in figure~\ref{fig:mapwavdestarch}. The top figure corresponds to the
destriped map after one iteration of the algorithm. Residual stripes dominate the right top corner
of the map. These are significantly reduced by the second iteration at shown in the bottom plot. 
It is important to notice that destriping allows us to recover the diffuse Galactic emission
on the Gemini region (middle left of the figure). \\

The wavelet destriping algorithm presented here is mainly based on the minimization of the variance
per pixel in the final map. As discussed in  \cite{thesealexandre} and \cite{arch_pipe} this approach
is not optimal for the Archeops data because of the atmospheric emission. A better approach is to
minimize in the map the ratio between the variance perpendicular and parallel to the scanning direction,
$\frac{\sigma_{\parallel}^2}{\sigma_{\perp}^2}$. The generalization of the wavelet destriping algorithm
to this latter approach is straight forward. As its implementation is much harder than when
considering Gabor atoms as base functions, it has not being considered for the analysis of the
Archeops data. However, despite of a simpler approach the quality wavelet destriped maps can
be compared to that of the final destriped maps used for the Archeops analysis (\cite{arch_pipe}).  

\subsection{Wavelet detrending of the Archeops data}

\begin{figure}[t]
\caption{{\it Top}: Fourier filtered map of the Archeops 353~GHz data {\it Bottom:} Wavelet filtered
map of the of the Archeops 353~GHz data. See text for details.}
\label{fig:wavdetrendingmap}
\end{figure}

As shown above, the Archeops destriped maps contain residual stripes which need to be removed
before any scientific analysis of the data. In general these stripes are superposed to the 
sky signal of interest and therefore a careful filtering is needed. Further, as the Galactic signal
in the Archeops maps is spike-like any filtering will produce ringing in the final maps. 
To overcome these problems, we have developed a wavelet based detrending algorithm.
First of all, we mask the Galactic signal and obtain a first approximation of
the low frequency components of the data by fitting them to a base of Gabor atoms.
Then, we use the fitted data to interpolate over the Galactic mask. Finally,
via a wavelet denoising algorithm we obtain the trends on the interpolated data which are then removed
before map making. For denoising we use a wavelet thresholding algorithm limited to
the few first smaller scales, typically up to $j=9$ when working with 6-millions data
samples. \\

Figure~\ref{fig:wavdetrendingtimeline} shows the performance of the detrending algorithm in
time domain. We plot in orange the interpolation to the data obtained by fitting Gabor atoms
and in blue the baseline obtained via the wavelet algorithm. We clearly observe that the 
the wavelet baseline manages to follow much better the accidents on the data reducing both
the residual stripes and ringing in the final maps. The bottom panel of figure~\ref{fig:wavdetrendingmap}
represent the wavelet filtered map at 353~GHz which can be compared to the Fourier filtered map in the
top panel of the figure. In the wavelet filtered maps the residual stripes and ringing are much
less. Further the diffuse structure of the Galactic emission is better preserved. Notice that 
the wavelet filtered maps at 353~GHz were used for the determination of the polarization of
the diffuse Galactic dust emission with Archeops (\cite{arch_polar}).

\section{Conclusions}
\label{conclusions}

The discrete wavelet transform (DWT), maximum overlap discrete wavelet transform (MODWT) and the
discrete wavelet packet transform (DWPT), because of their simultaneous time and frequency localization
properties, are very important tools for the analysis of TOD from CMB experiments.
They allow us to trace the evolution in time of the data power spectrum and to easily identify
systematic effects on the data. \\

The DWT permits a compact representation of Gaussian stationary noise. Indeed, it decorrelates
Gaussian $1/f$-type noise which is commonly present in the detector of CMB experiments. Within
the wavelet description the covariance matrix of Gaussian $1/f$-type noise is diagonal as it is
the case in the Fourier space. This allows us to efficiently and accurately simulate Gaussian $1/f$-type noise
using wavelets. Moreover, the DWT transform permits a straight forward description of locally
stationary Gaussian noise via the time evolution of the variance of the
wavelet coefficient but with a diagonal covariance matrix. A particular case of this is the 
time modulated Gaussian noise for which the time evolution is common to all the wavelet scales. \\
  
The above properties allows us to generalize the Fourier space algorithms for fast optimal
map making and maximum likelihood determination of the CMB angular power spectrum to the
wavelet space including both stationary and locally stationary Gaussian noise. At this respect,
we have developed a wavelet based destriped algorithm which both in simulated and true
Archeops data reduce significantly the level of stripes in the maps. Despite the fact that this
algorithm is based on a simple approach of minimization of the variance per pixel, the results
obtained can be compared to those from more precise destriping algorithms. \\

Finally we have performed a full wavelet analysis of the Archeops data. The visualization and
careful study of the time-frequency space via the DWPT allows us to clearly identify systematics
and to characterize the quality of the data. Further, we have proceeded to the careful wavelet modeling
of the locally stationary noise in the Archeops data. This modeling allowed us to obtain 
via simulations the angular power spectrum of the Archeops noise needed for the estimation
of the CMB angular power spectrum with Archeops (\cite{benoit_cl}). We have applied the wavelet
destriping algorithm to the Archeops data obtaining encouraging results. Finally, we have
developed a detrending algorithm based on a wavelet denoising of the data. This algorithm
applied to the Archeops destriped data reduce significantly the residual stripes on the
final maps and introduces very little ringing. The wavelet detrended Archeops maps at
353~GHz were used for the determination of the polarized diffuse emission from Galactic
dust (\cite{arch_polar}).  

\begin{acknowledgements}
Fist of all we would like to thank the Archeops collaboration for allowing us to analyse the
Archeops data and to present the results in here.
We also thank B. Vidakovic, D.B. Percival and X. Desert for very useful discussion during the
writting of this paper.
\end{acknowledgements}


\begin{thebibliography}{}

\bibitem[Bennett et al. {2003}]{wmap}
Bennett C.L. et al., 2003, \apj, 148, 1

\bibitem[Beno{\^\i}t~{\it et~al}~2002]{benoit_app}
Beno{\^\i}t, A., Ade, P., Amblard, A.,~{\it et~al}, 2002,
Astropart. Phys. 17, 101

\bibitem[Beno{\^\i}t~{\it et~al}~2003a]{benoit_cl}
Beno{\^\i}t, A., Ade, P., Amblard, A.,~{\it et~al} 2003a, A\&A. 399, L25

\bibitem[Beno{{\^\i}}t~{\it et~al}~2003b]{benoit_params}
Beno{{\^\i}}t, A., Ade, P., Amblard, A.,~{\it et al} 2003b, A\&A. 399, L19

\bibitem[Beno\^\i t {\it et al}~2004]{arch_polar}
Beno{\^\i}t, A., Ade, P., Amblard, A.,~{\it et~al}, 2004, A\&A, 424, 571

\bibitem[Borrill (1999)]{madcap} Borrill, J., 1999, In {\it Proceedings of the
    5th European SGI/Cray MPP Workshop\/}, Bologna, Italy, astro-ph/9911389

\bibitem[Bourrachot, A. 2004]{thesealexandre}
Bourrachot, A., 2004, PhD Thesis, Universit\'e Paris XI, UFR scientifique d'Orsay
  
\bibitem[{Coifman, R.R. \& Wickerhauser, M.V. 1992}]{entbestbasis}
 Coifman, R.R. \&  Wickerhauser, M.V., 1992, IEEE Transactions on Information Theory, {\bf 38}, 713

\bibitem[{Daubechies,I. 1992}]{daubbook}
 Daubechies, I., 1992, Ten Lectures on wavelets, Philadelphia: SIAM

\bibitem[{Dickinson {\it et al.} 2004}]{vsapaper}
 Dickinson, C. {\it et al.}, 2004, MNRAS, 353, 732

\bibitem[{Dor\'e {\it et al.} 2001}]{mapcumba}
 Dor\'e, O. {\it et al.}, 2001, A\&A, {\bf 374}, 538

\bibitem[{Efstathiou, G} 2005]{efstathioudest}
Efstathiou, G., 2005, \mnras, 356, 1549-1558 

\bibitem[Finkbeiner~{\it et~al}~1999]{fds}
Finkbeiner, D., Davis, M., Schlegel, D., 1999, ApJ, 524, 867

\bibitem[{Fry\'zlewicz, P., Van Bellegem, S. \& von Sachs, R. 2002}]{timemodnoisepaper}
 Fry\'zlewicz, P., Van Bellegem, S. \& von Sachs, R., 2002, Universit\'e Catholique de Louvain,
 discussion paper 0208

\bibitem[{Halverson {\it et al.}  2002}]{dasipaper}
 Halverson, N.W. {\it et al.}, 2002, MNRAS, {\bf 586}, 38

\bibitem[{Hivon {\it et al.} 2002}]{master}
 Hivon, E. {\it et al.}, 2002, ApJ, {\bf 567}, 2

\bibitem[{Keih{\" a}nen, E. } {\it et al} 2004]{keihanendest}
Keih{\" a}nen, E., Kurki-Suonio, H., Poutanen, T., Maino, D. \& Burigana, C., 2004, \aap, 428, 287-298

\bibitem[Kogut~{\it et~al}~2003]{wmappol}
Kogut, A., Spergel, D.~N., Barnes, C., 2003, ApJ Suppl. 148, 161

\bibitem[{Lai, M-J. 1995}]{laipaper}
 Lai, M.-J., 1995, IEEE Transactions on Signal Processing, 43, 2203-5

\bibitem[{Lee {\it et al}. 2001}]{maxpaper}
 Lee, A.T. {\it et al}. 2001, ApJ, {\bf 561}, L1

\bibitem[{Mac\'{\i}as-P\'erez}, J.F.  {\it et al}. 2005]{arch_pipe}
Mac\'{\i}as-P\'erez, J.F., Bourrachot, A.  {\it et al}.,  2005, \aap, in preparation

\bibitem[{Maino, D.  {\it et al}. 2002}]{mainodest}
Maino, D., Burigana, C., G{\' o}rski, K.~M.,Mandolesi, N., \& Bersanelli, M., 2002, \aap, 387, 356-365

\bibitem[{Mallat, 1989}]{mallatpyramid} 
 Mallat, S.G., 1989, IEEE Transactions on Pattern Analysis and Machine intelligence, {\bf 11}, 674-693

\bibitem[{Mallat \& Zhang, Z., 1993}]{matchingpursuitref}
 Mallat, S.G.\& Zhang, Z. 1993, IEEE Transactions on Signal Processing, {\bf 41(12)}, 3397-3415

\bibitem[{Mallat {\it et al}. 1998}]{adaptivecovest}
 Mallat, S.G., Papanicolau, G. \& Zhang, Z. 1998, Annals of Statistics, {\bf 26(1)}, 1-47

\bibitem[{Natoli {\it et al.} 2001}]{mapmakingitalian}
 Natoli, P., de Gasperis, G., Gheller, C. \& Vittorio, N., 2001, A\&A, {\bf 372},346-356 

\bibitem[{Netterfield {\it et al.} 2002}]{boompaper}
 Netterfield, C.B. {\it et al.}, ApJ, {\bf 571}, 604

\bibitem[{Pearson {\it et al.}} 2002]{cbipaper}
 Pearson, T.J. {\it et al.}, 2002, AAS, 34, 949

\bibitem[{Percival \& Walden 2000}]{percibook}
 Percival D.B. \& Walden A.T., 2000, Wavelet Methods for time series analysis, 
 Cambridge University Press

\bibitem[{Percival {\it et al.} 2000}]{wavestrapping}
 Percival D.B., Sardy, S. \& Davison A. C., 2000, Wavestrapping Time Series: Adaptive
 Wavelet-based Bootstrapping in Non-Linear and non-stationary Signal Processing,
 Cambridge University Press.

\bibitem[{Poutanen, T.} {\it et al.} 2004]{poutanendest}
Poutanen, T., Maino, D., Kurki-Suonio, H., Keih{\" a}nen, E. \& Hivon, E., 2004, \mnras, 353, 53-58 

\bibitem[{Press, W.H.} {\it et al.} 1996]{numrec}
Press, W.H., Vetterling, W.T., Teukolsky, S.A. \& Flannery, B.P., 1996, Numerical Recipes in C.

\bibitem[{Revenu, B.} {\it et al.} 2000]{revenudest}
Revenu, B., Kim, A., Ansari, R., Couchot, F.Delabrouille, J. \& Kaplan, J., 2000 , \aap, 142, 499-509

\bibitem[{Sbarra, C.} {\it et al.} 2003]{sbarradest}
 Sbarra, C., Carretti, E., Cortiglioni, S., Zannoni, M., Fabbri, R., Macculi, C. \& Tucci, M., 2003, \aap, 401, 1215-1222

\bibitem[{Smoot {\it et al. 1992}}]{cobepaper}
 Smoot, G.F. {\it et al.}, 1992, ApJ, {\bf 396}, L1-L5

\bibitem[{Tristram {\it et al.} 2005}]{tristram_cl}
 Tristram, M., Patanchon, G.,  Mac\'{\i}as-P\'erez, J.F.  {\it et al.}, 2005, \aap, 436, 785

\bibitem[{Vidakovic,B. 1999}]{vidakovicbook}
 Vidakovic, B., 1999, Statistical modeling by Wavelets.

\bibitem[{Wada S. and Ito N., 2000}]{Wavelet:01} 
 Wada S. and Ito N., DSP 2000 Workshop

\bibitem[{Yvon {\it et al. 2005}}]{mirage}
 Yvon, D. {\it et al.}, 2005, A\&A, 436, 729


\end{thebibliography}
\end{document}